\documentclass[aip,pop,preprint,showpacs,groupedaddress,amsmath,amssymb]{revtex4-1}
\usepackage[usenames,dvipsnames]{color}

\usepackage{graphicx}

\newcommand{\bew}{\begin{widetext}}
\newcommand{\eew}{\end{widetext}}
\newcommand{\beq}{\begin{equation}}
\newcommand{\eeq}{\end{equation}}
\newcommand{\bea}{\begin{eqnarray}}
\newcommand{\eea}{\end{eqnarray}}
\newcommand{\bes}{\begin{subequations} \bea}
\newcommand{\ees}{\eea \end{subequations}}
\newcommand{\mbf}[1]{\mathbf{#1}}

\newcommand{\mrm}[1]{\mathrm{#1}}
\newcommand{\del}{\nabla}
\newcommand{\divr}{\nabla \cdot}

\newcommand{\epsi}{\varepsilon}
\newcommand{\dsub}[1]{\partial_{#1}}
\newcommand{\wt}[1]{\widetilde{#1}}

\newcommand{\abs}[1]{{\vert {#1} \vert}}

\newcommand{\mean}[2]{{\langle {#1} \rangle}_{#2}}

\begin{document}

%%%%  RESET LINE NUMBER FONT
%%%%  AND MARGIN SPACING
\makeatletter
\@ifundefined{linenumbers}{}{
\renewcommand\linenumberfont{\normalfont\scriptsize}
\setlength\linenumbersep{0.5cm}
}
\makeatother

%\preprint{}

\title{Remarks on the derivation and evaluation of the Stacey-Sigmar model for tokamak equilibrium} %Title of paper



\author{Robert W. Johnson}
\email[]{robjohnson@alphawaveresearch.com}
\homepage[]{http://www.alphawaveresearch.com}
%\thanks{}
\affiliation{Alphawave Research, Jonesboro, GA 30238, USA}


\date{\today}

\begin{abstract}
The Stacey-Sigmar model for tokamak equilibrium as presented in the literature relies heavily on the neoclassical theory of the electrostatic field.  Its neglect of Gauss's law is inconsistent with the potential formulation of electrodynamics.  Its treatment of the dynamic electric field generated by the central heating and poloidal field coils also is suspect.  Its derivation of the viscosity term remains incomplete and does not account for the varying pitch angle of the magnetic field.  A derivation of the viscosity term which respects the pitch angle can account for the radial force commonly ascribed to the radial electrostatic field, thereby obviating the desire to neglect Gauss's law.
\end{abstract}


\maketitle %\maketitle must follow title, authors, abstract and \pacs


In their recent paper~\cite{bae-043011}, Bae, Stacey, and Solomon present a model for plasma rotation within a tokamak.  Their work focuses on the adaptation of the concentric circular flux surface model~\cite{frc-pop-2006} to a geometry which better represents what is believed to exist within a D shaped tokamak~\cite{stacey-082501}.   Part of that belief system is the assertion that a neutral fluid model can support the presence of an electrostatic field.  To accomplish that presumption, the neoclassical interpretation of the quasi-neutral approximation states that Gauss's law is not to be considered among the fundamental equations governing the physics.  Such an approach is not consistent with the potential formulation of electrodynamics, however, as it ascribes an incorrect number of degrees of freedom to the electromagnetic fields.

The authors assume in their Eqn.~(41) a form for the electrostatic potential of \beq
\Phi_\mathrm{model}(r,\theta) \equiv \bar{\Phi}(r) [1 + \Phi^c(r) \cos (\theta) + \Phi^s(r) \sin (\theta)] \; ,
\eeq where the approximation is between the physical (unknowable) potential and that of the model $\Phi_\mathrm{physical} \approx \Phi_\mathrm{model}$, not between the symbol for the model's potential and its definition in terms of the Fourier degrees of freedom considered by the model.  The same form is assumed for the electron density in Eqn.~(22a) while the electron temperature is assumed to be constant over the flux surface, thus the model for the electron pressure is defined as \beq
p_e(r,\theta) \equiv T_e(r) \bar{n}_e(r) [1 + n_e^c(r) \cos (\theta) + n_e^s(r) \sin (\theta)] \; .
\eeq  Their expression in Eqn.~(19) for the physical electron poloidal equation of motion in units of force density retains only the pressure and electric field terms \beq \label{eqn:physpol}
0 = h_\theta^{-1} \dsub{\theta} p_e(r,\theta) + e n_e(r,\theta) E_\theta(r,\theta) \; ,
\eeq for $E_\theta = - h_\theta^{-1} \dsub{\theta} \Phi$ in the static case, which can be written in terms of the degrees of freedom specified by the Stacey-Sigmar model as \bea \label{eqn:polfore}
0 &=& - e \bar{\Phi} [1 + n_e^c \cos (\theta) + n_e^s \sin (\theta)] [\Phi^s \cos (\theta) - \Phi^c \sin (\theta)] \nonumber \\
 & & + T_e [n_e^s \cos (\theta) - n_e^c \sin (\theta)] \; ,
\eea where the poloidal measure factor $h_\theta$ and the mean electron density $\bar{n}_e$ cancel out of the equation and the dependence on $r$ is implicit; note the distinction between Eqn.~(\ref{eqn:physpol}) which represents the physics and Eqn.~(\ref{eqn:polfore}) which defines the Stacey-Sigmar model thereof.  The authors then claim in their Eqn.~(42) that the approximate solution \beq \label{eqn:CSSsoln}
\left[ \begin{array}{c} \Phi^c \\ \Phi^s \end{array} \right] \approx \left[ \begin{array}{c} n_e^c \\ n_e^s \end{array} \right] T_e / e \bar{\Phi}
\eeq is valid over the entire region of consideration.  Let us now examine under what conditions the RHS expression constitutes a valid solution.

To account for the toroidal geometry of the device, one must integrate the force density over the flux surface to evaluate the net force on the plasma medium, as is done for the ion equations of motion.  Considering first the concentric circular flux surface approximation~\cite{frc-pop-2006} upon which their model~\cite{bae-043011} is based, the unity, cosine, and sine moments of the flux surface average are given by the integral expressions \beq
\left[ \begin{array}{c} X_U (r) \\ X_C (r) \\ X_S (r) \end{array} \right] = (2 \pi)^{-1} \int_{-\pi}^{\pi} d\theta [1 + \epsi \cos (\theta)] X(r,\theta) \left[ \begin{array}{c} 1 \\ \cos (\theta) \\ \sin (\theta) \end{array} \right] \; ,
\eeq where $\epsi \equiv r / R_0$ is the ratio of the minor radial location to the central major radius.  Taking those moments of Eqn.~(\ref{eqn:polfore}) yields the system \bes \label{eqn:polsystem}
\left[ \begin{array}{c} 0 \\ 0 \\ 0 \end{array} \right] &\propto& \left[ \begin{array}{c} 
4 e \bar{\Phi} \left( n_e^s \Phi^c - n_e^c \Phi^s\right) + 4 \epsi ( T_e n_e^s - e \bar{\Phi} \Phi^s ) \\
4 ( T_e n_e^s - e \bar{\Phi} \Phi^s ) + \epsi e \bar{\Phi} ( n_e^s \Phi^c-3 n_e^c \Phi^s ) \\
4 ( e \bar{\Phi} \Phi^c - T_e n_e^c ) + \epsi e \bar{\Phi} ( n_e^c \Phi^c-n_e^s \Phi^s ) 
\end{array} \right] \\
 &\equiv& \left[ \begin{array}{c} U \\ C \\ S \end{array} \right] \; ,
\ees where the constant of proportionality is equal to 1/8.  The terms with a factor of $\epsi$ arise explicitly from consideration of the toroidal measure factor $R / R_0 = [1 + \epsi \cos (\theta)]$ and would not be present for a cylindrical containment vessel $R_0 \rightarrow \infty$ with vanishing aspect ratio.  In the limit $\epsi \rightarrow 0$ one can indeed say that Eqn.~(\ref{eqn:CSSsoln}) gives a solution of equations $C$ and $S$ with $U$ satisfied identically; however, in that limit $n_e^{c,s} \rightarrow 0$ by continuity.  The proposed solution in Eqn.~(\ref{eqn:CSSsoln}) is in fact a zeroth order expression equivalent to the statement $0 = 0$, a fact which the authors obfuscate by introducing the arbitrary renormalization $n_e^{c,s} \equiv \epsi \wt{n}_e^{c,s}$.  For finite $R_0$, Eqn.~(\ref{eqn:CSSsoln}) is valid only on the magnetic axis $r = 0$, yet the authors apply this model throughout the confinement region where, at its greatest, the flux surface aspect ratio $\epsi$ approaches 1/3.  Passing the full system of equations $[U,C,S]$ to one's favorite computer algebra software, one finds that no explicit solution can be found for $[\bar{\Phi}, \Phi^c, \Phi^s]$.  The expression for $U$ is conspicuous by its absence from the model~\cite{bae-043011,frc-pop-2006}, when the unity moment of every other equation is considered.

One can linearize the form of Eqn.~(\ref{eqn:polsystem}) by rewriting $[U,C,S]$ in terms of $T_e / e \bar{\Phi}$, yielding the equivalent system \beq
\left[ \begin{array}{c} 0 \\ 0 \\ 0 \end{array} \right] \propto 
\left[ \begin{array}{ccc}
 4 \epsi n_e^s & 4 n_e^s & - 4 ( n_e^c + \epsi ) \\
 4 n_e^s & \epsi n_e^s & - ( 4 + 3 \epsi n_e^c ) \\
 -4 n_e^c & ( 4 + \epsi n_e^c ) & - \epsi n_e^s
\end{array} \right] 
\left[ \begin{array}{c} T_e / e \bar{\Phi} \\ \Phi^c \\ \Phi^s \end{array} \right] \; .
\eeq  In matrix form, one should instantly recognize a set of linear, homogeneous algebraic equations whose only (unique) solution is trivial, $[T_e / e \bar{\Phi}, \Phi^c, \Phi^s] = [0, 0, 0]$.  If one asserts that $T_e$ is not zero, then $\bar{\Phi}$, defined as the mean potential difference between locations on the flux surface and the magnetic axis, must be infinite.  Since the secondary authors (Stacey and Solomon) were informed of this issue~\cite{rwj-ttf12} as early as 2007, their continued support for the use of a model with such obvious difficulties is hard to understand.

A detailed investigation of these equations published elsewhere~\cite{rwj-mrc02} considers the expansion to third order of the logarithmic expression  \bes
0 &=& T_e(r) \dsub{\theta} n_e(r,\theta) - e n_e(r,\theta) \dsub{\theta} \Phi(r,\theta) \\
 &=& [\dsub{\theta} n_e(r,\theta)] / n_e(r,\theta) - \dsub{\theta} e \Phi(r,\theta) / T_e(r) \\
 &=& \dsub{\theta} \lbrace \log [n_e(r,\theta) / C(r)] - e \Phi(r,\theta) / T_e(r) \rbrace \; .
\ees  Inserting the degrees of freedom specified by the Stacey-Sigmar model into that expression simply recovers Eqn.~(\ref{eqn:polfore}), which can be satisfied only by the trivial solution in toroidal geometry.  According to the theory by Maxwell, the electrostatic field (more specifically the scalar potential in Lorenz gauge) is independent of the apparent motion of the sources, thus it must be determined by the Poisson equation and not an equation of motion.

%  One can achieve the same result without recourse to the flux surface average simply by evaluating Eqn.~(\ref{eqn:polfore}) at any three non-degenerate angles $[\theta_1,\theta_2,\theta_3]$.  The resulting equations $E_i$ for $i \in \{1,2,3\}$, given explicitly by \bea
%0 \propto E_i &\equiv& 2 [ n_e^c \sin (\theta_i) - n_e^s \cos (\theta_i) ]  T_e / e \bar{\Phi} \nonumber \\
% & & - [ n_e^c \sin ( 2 \theta_i) - n_e^s \cos ( 2 \theta_i) + 2 \sin (\theta_i) + n_e^s ] \Phi^c \nonumber \\
% & & + [ n_e^s \sin ( 2 \theta_i) + n_e^c \cos ( 2 \theta_i) + 2 \cos (\theta_i) + n_e^c ] \Phi^s \; ,
%\eea will always be linearly independent on account of the trigonometric functions.

Some further, ancillary remarks now follow.  Since the integrals over the Miller flux surfaces have to be done numerically, one wonders why the authors do not work directly with the 2-D equilibrium data provided by the experimentalists rather than its 1-D summary.  A derivation of their new term $\nu_{\mathrm{d} j}^1$ found in Eqn.~(34), for the Stacey-Sigmar model in the concentric circular flux surface approximation, has been available in the literature~\cite{rwj-jpp05} since 2011 (and earlier on the arXiv); a term arising from the change in the gyroviscous coefficient dependent upon the gyrofrequency still is missing from their $\nu_{\mathrm{d} j}^2$.  One should note that the effect of $\nu_{\mathrm{d} j}^1$ points in the direction opposite to that of $\nu_{\mathrm{d} j}^2$; in other words, the new Stacey-Sigmar model~\cite{bae-043011} claims the gyroviscous force pushes in a direction opposite to that claimed for the last 30 years~\cite{stacandsig-1985}.  In their evaluation of the unity moment of the radial momentum balance, Eqn.~(B4), they neglect the contributions of the inertial term $-\mean{\hat{\mbf{r}} \cdot n m (\mbf{V} \cdot \del) \mbf{V}}{}$, and of the radial shear viscous force $-\mean{\hat{\mbf{r}} \cdot \divr \Pi_S}{}$ on account of their assumption in Eqn.~(A7) that $\abs{B_\theta / B} \approx 0$, which together are sufficient to balance the force from the pressure gradient and $\mbf{V} \times \mbf{B}$ terms~\cite{rwj-jpp05} without invoking the presence of a radial electric field $E_r$.

Finally, the authors in several places mention comparison of their model~\cite{bae-043011} with results obtained in the concentric circular flux surface approximation~\cite{frc-pop-2006}.  Readers should be made aware that the calculations presented in that paper did not include the effect of the toroidal electric field despite its Eqn.~(27) implying the contrary.  The relevant lines of the code used for that evaluation read:
%\bew
\begin{verbatim}
202:      Y_i = extmomhat_tor_i + beta*vth_i*vr_hat_i/nustar_iz
203:      Y_z = extmomhat_tor_z + beta*vth_z*vr_hat_z/nustar_zi
\end{verbatim}
%\eew 
which account for the NBI momentum input and the radial flux term but not the toroidal electric field.  Whether or not the authors~\cite{bae-043011} are accounting for $E_\phi$ numerically is unclear, since the only experimental profiles displayed are those of the poloidal and toroidal components of the carbon velocity.  The presence of the loop voltage in its Table 1 suggests that it is; however, correspondence with the primary author~\cite{bae-words} indicates that such may not be the case.  In order to assess the reproducibility of their results, the authors should clearly state how the toroidal electric field is evaluated in their numerical calculation.  Unpublished investigations (available on the arXiv) indicate that its effect on the toroidal velocities is not insignificant, and a method for its determination given the time rate of change of the current through the central solenoid and poloidal field shaping coils is available in the literature~\cite{rwj-epjd01}.

In closing, a few remarks on the source of the difficulties faced by the neoclassical model for plasma physics are in order.  Much confusion abounds over what many call the quasi-neutral approximation, which might be better nominated as the neutral fluid limit.  For comparison, the quasi-static approximation states that $\divr \mbf{J}_\mathrm{model} = 0$ when $\divr \mbf{J}_\mathrm{physical} = - \dsub{t} j_\mathrm{physical}$ is vanishingly small; its electrostatic analogue is $\divr \epsilon_0 \mbf{E}_\mathrm{model} = 0$ when $\divr \epsilon_0 \mbf{E}_\mathrm{physical} = j_\mathrm{physical}$ is vanishingly small, such that $\mbf{E}_\mathrm{physical} \approx \mbf{E}_\mathrm{model} = 0$.  No theory of electromagnetism is complete without the inclusion of Gauss's law, which is conspicuous by its absence from the model equations~\cite{bae-043011,frc-pop-2006}.  For example, consider Chapters~2 and 4 in Dendy's textbook~\cite{dendybook-93}.  In Chapter~2, Elliott specifies the low frequency MHD system in terms of 14 scalar equations, retaining only the curl equations from the Maxwell system, and variables $(\rho,p,\mbf{V},\mbf{J},\mbf{E},\mbf{B})$ representing mass density, pressure, fluid velocity, current density, and the electromagnetic fields; however, in Chapter~4, Hopcraft includes explicitly a 15th equation $\divr \mbf{B} = 0$ (sometimes known as Gauss's law for magnetism) which no one would argue against unless they are holding onto a magnetic monopole.  To reach a count of 14 degrees of freedom, one must ascribe a total of 6 degrees of freedom to the electromagnetic fields, contrary to the physical requirement of 3 degrees of freedom in matter and 2 in vacuum, which are identified with the polarization states of the constituent photons.

As every student of electromagnetic field theory should know, the classical equations of Maxwell can be expressed most succinctly~\cite{davis70,ryder-qft,naka-798212} as ${\mrm{d} \star \mrm{d} A = J}$ for ${\mrm{d} \mrm{d} A \equiv 0}$, in terms of the exterior derivative $\mrm{d}$, the Hodge dual $\star$, the connection 1-form $A$, and the current 3-form $J$.  Those equations in coordinate-free notation, while impractical for any particular calculation, demonstrate the physical equivalence of the inhomogeneous Maxwell equations; the Maxwell-Ampere relation is nothing but Gauss's law in a different frame of reference.  The homogeneous equations are satisfied identically in the potential formulation, while the inhomogeneous equations are constrained by continuity of the source $\dsub{\mu} J^\mu = 0$ and the potential $\dsub{\mu} A^\mu = 0$.  One can easily show that even the most basic of derivations, such as that for the cold plasma dispersion relation~\cite{rwj-pop02}, when done in the field formulation without Gauss's law, are not consistent with results derived from the potential formulation of electrodynamics.  Simply put, no one can do better than Maxwell by doing less.

%\begin{acknowledgments}
% Put your acknowledgments here.
%\end{acknowledgments}

%%%%  END LINE NUMBERS
\makeatletter
\@ifundefined{linenumbers}{}{
\nolinenumbers
}
\makeatother

%\newpage

\bibliography{../plasma.bib,../particle.bib}

%%% RESPONSE
%\newpage
%\input{response04}

\end{document}